\documentclass[10pt,english]{article}
\usepackage[T1]{fontenc}
\usepackage[latin1]{inputenc}

\newcommand{\noun}[1]{\textsc{#1}}

\usepackage{slashed}

\usepackage{babel}
\makeatother

\usepackage{babel}

\begin{document}

\title{\textbf{Implementing inverse seesaw mechanism in }$SU(3)_{c}\otimes SU(4)_{L}\otimes U(1)_{X}$\textbf{
gauge models }}

\author{\noun{ADRIAN} PALCU}

\date{\emph{Faculty of Exact Sciences - ''Aurel Vlaicu'' University of
Arad, Str. Elena Dr\u{a}goi 2, 310330 - Arad, Romania}}
\maketitle
\begin{abstract}
Generating appropriate tiny neutrino masses via inverse seesaw mechanism
within the framework of a particular $SU(3)_{c}\otimes SU(4)_{L}\otimes U(1)_{X}$
gauge model is the main outcome of this letter. It is achieved by
simply adding three singlet exotic Majorana neutrinos to the usual
ones included in the three lepton quadruplet representations. The
theoretical device of treating gauge models with high symmetries is
the general method by Cot\u{a}escu. It provides us with a unique
free parameter ($a$) to be tuned in order to get a realistic mass
spectrum for the gauge bosons and charged fermions in the model. The
overall breaking scale can be set around 1-10 TeV so its phenomenology
is quite testable at present facilities.


\end{abstract}

\section{Introduction}

It is well-known that the Standard Model (SM) (\cite{key-1} - \cite{key-3})
- based on the gauge group $SU(3)_{c}\otimes SU(2)_{L}\otimes U(1)_{Y}$
undergoing a spontaneous symmetry breaking (SSB) in its electro-weak
sector up to the universal $U(1)_{em}$ - is not a sufficient device,
at least for some stringent issues in the particle physics today.
When it comes to generating neutrino tiny masses \cite{key-4,key-5},
the framework of the SM is lacking the needed ingredients, so one
should call for some extra considerations which are less natural in
the context. One of the ways out seems to be the enlargement of the
gauge group of the theory as to include naturally among its fermion
representations some right-handed neutrinos - mandatory.elements for
some plausible mass terms in the neutrino sector Yukawa Lagrangian
density (Ld).

Among such possible extensions of the SM, the so called ''3-3-1''
and {}``3-4-1'' classes of models- where the new electroweak gauge
groups are $SU(3)_{L}\otimes U(1)_{X}$ \cite{key-6} - \cite{key-8}
and $SU(4)_{L}\otimes U(1)_{X}$ \cite{key-9} - \cite{key-14} respectively
- has meanwhile established themselves as much suitable candidates.
Some systematic classifications \cite{key-15} - \cite{key-19} of
these SM-extensions have been done. In this paper we are concerned
with a particular class of 3-4-1 models (namely the one that prohibits
exotic electric charges) whose phenomenological analysis can be found
in the literature, see Refs. \cite{key-20} - \cite{key-28} The neutrino
mass issue has been addressed \cite{key-29} - \cite{key-31} with
viable results within the framework of such models.

Here we propose a slightly different approach from the canonical one,
in the sense that we apply the prescriptions of the general method
\cite{key-32} of treating gauge models with high symmetries. Proposed
initially by Cot\emph{\u{a}}escu, it essentially consists of a general
algebraical procedure in which electro-weak gauge models with high
symmetries ($SU(N)_{L}\otimes U(1)_{Y}$) achieve their SSB in only
one step up to the residual $U(1)_{em}$ by means of a special Higgs
mechanism. The scalar sector is organized as a complex vector space
where a real scalar field $\varphi$ is introduced as the norm for
the scalar product among scalar multiplets. It also ensures the orthogonality
in the scalar vector space. Thus, the survival of some unwanted Goldstone
bosons is avoided. This leads to a one-parameter mass spectrum, due
to a restricting trace condition that has to hold throughout. The
compatibility of this particular method with the canonical approach
to 3-3-1 and 3-4-1 models in the literature was proved in some recent
papers by the author\cite{key-33} - \cite{key-37}. In the case of
the particular 3-3-1 models with right-handed neutrinos an appealing
outcome \cite{key-37} with only two physical massive Higgses with
non-zero interactions finally emerged. 

Once we established the framework in which the 3-4-1 gauge model of
interest is treated, we exploit the realization of a kind of quasi-inverse
seesaw mechanism \cite{key-38} - \cite{key-46} by simply adding
3 new exotic sterile Majorana singlets $\left(N_{R}\right)$. Finally,
the free parameter (let's call it $a$) is tuned in order to obtain
the whole mass spectrum (including the neutrinos). An apparently unused
up to now parameter $\eta_{0}$ in the general method proves itself
here as the much needed ''lepton number violating'' coupling to
achieve the Majorana mass terms for $N_{R}$ in the neutrino sector.

The letter is organized as follows. It begins with a brief presentation
of the model and its parametrization supplied by the general Cot\u{a}escu
method (in Sec.2) and continues with the inverse seesaw mechanism
worked out within this framework (Sec. 3) and the tuning of the parameters
(Sec. 4) in order to obtain phenomenologically viable results for
the neutrino masses. Some conclusions are sketched in the last section
(Sec. 5).

\section{The 3-4-1 gauge model}

Let's start by presenting the anomaly-free particle content of the
3-4-1 gauge model of interest here. It comprises the following:

\textbf{Lepton families}

\begin{equation}
L_{\alpha}=\left(\begin{array}{c}
e_{\alpha}\\
\nu_{\alpha}\\
N_{\alpha}^{\prime}\\
N_{\alpha}^{\prime\prime}\end{array}\right)_{L}\sim(\mathbf{1,4^{*}},-1/2)\quad\quad\quad e_{\alpha R}\sim(\mathbf{1,1},-2)\label{Eq.1}\end{equation}
with $\alpha=1,2,3$.

\textbf{Quark families}\begin{equation}
\begin{array}{ccc}
Q_{iL}=\left(\begin{array}{c}
u_{i}\\
-d_{i}\\
D_{i}\\
D_{i}^{\prime}\end{array}\right)_{L}\sim(\mathbf{3,4},-1/6) &  & Q_{3L}=\left(\begin{array}{c}
d_{3}\\
u_{3}\\
U\\
U^{\prime}\end{array}\right)_{L}\sim(\mathbf{3},\mathbf{4^{*}},5/6)\end{array}\label{Eq.2}\end{equation}
\begin{equation}
\begin{array}{c}
d_{3R},d_{iR},D_{iR},D_{iR}^{\prime}\sim(\mathbf{3},\mathbf{1},-2/3)\end{array}\label{Eq.3}\end{equation}

\begin{equation}
u_{3R},u_{iR},U_{R},U_{L}^{\prime}\sim(\mathbf{3},\mathbf{1},4/3)\label{Eq.4}\end{equation}
with $i=1,2$.

The above representations (written in the usual notation) ensure the
cancellation of all the axial anomalies by an interplay between families.
This prevents the model from compromising its renormalizability by
triangle diagrams. The capital letters are reserved for the exotic
quarks ($D_{i}$, $D_{i}^{\prime}$ and $U$, $U^{\prime}$) in each
family. They must be heavier than the ordinary quarks known from the
SM in order to keep consistency with the low energy weak phenomenology.

To this fermion content one can add 3 Majorana exotic neutrinos $N_{\alpha R}\sim(\mathbf{1},\mathbf{1},0)$
without the danger of spoiling the renormalizability. The advantage
these 3 exotic neutrinos bring is that they can play a crucial role
in realizing a particular sort of inverse seesaw mechanism\cite{key-38}
- \cite{key-46}.

\textbf{Gauge bosons}

The gauge bosons of the model are determined by the generators of
the associated $su(4)$ Lie algebra, expressed by the usual Gell-Mann
matrices $T_{a}=\lambda_{a}/2$ . So, the Hermitian diagonal generators
of the Cartan sub-algebra are in the fundamental representation:

\[
D_{1}=T_{3}=\frac{1}{2}{\textrm{Diag}}(0,1,-1,0)\,,\quad D_{2}=T_{8}=\frac{1}{2\sqrt{3}}\,{\textrm{Diag}}(0,1,1,-2)\]
\begin{equation}
D_{3}=T_{15}=\frac{1}{2\sqrt{6}}\,{\textrm{Diag}}(-3,1,1,1)\,.\label{Eq,5}\end{equation}
In order to discuss the phenomenology of this model, we employ the
Cot\u{a}escu method of treating gauge models with high symmetries.
For the sake of completeness we write down the electric charge operator
in this very method. It naturally arises as: $Q^{\rho}=e\left[-\sqrt{\frac{3}{2}}T_{15}^{\rho}+\frac{1}{2}X^{\rho}\right]$
for each representation $\rho$. Hence, one can easily recover the
above fermion representation (up to an unusual order in the quadruplets,
that can be rearranged at any time). 

In this basis the gauge fields are expressed by: $A_{\mu}^{0}$ (corresponding
to the Lie algebra of the group $U(1)_{X}$) and $A_{\mu}\in su(4)$,
that can be put as\begin{equation}
A_{\mu}=\frac{1}{2}\left(\begin{array}{ccccccc}
D_{\mu}^{1} &  & \sqrt{2}Y_{\mu} &  & \sqrt{2}X_{\mu} &  & \sqrt{2}X_{\mu}^{\prime}\\
\\\sqrt{2}Y_{\mu}^{*} &  & D_{\mu}^{2} &  & \sqrt{2}K_{\mu} &  & \sqrt{2}K_{\mu}^{\prime}\\
\\\sqrt{2}X_{\mu}^{*}{} &  & \sqrt{2}K_{\mu}^{*} &  & D_{\mu}^{3} &  & \sqrt{2}W_{\mu}\\
\\\sqrt{2}X_{\mu}^{\prime*} &  & \sqrt{2}K_{\mu}^{\prime*} &  & \sqrt{2}W_{\mu}^{*} &  & D_{\mu}^{4}\end{array}\right),\label{Eq.6}\end{equation}
with $D_{\mu}^{1}=A_{\mu}^{3}+A_{\mu}^{8}/\sqrt{3}+A_{\mu}^{15}/\sqrt{6}$,
$D_{\mu}^{2}=-A_{\mu}^{3}+A_{\mu}^{8}/\sqrt{3}+A_{\mu}^{15}/\sqrt{6}$,
$D_{\mu}^{3}=-2A_{\mu}^{8}/\sqrt{3}+A_{\mu}^{15}/\sqrt{6}$, $D_{\mu}^{4}=-3A_{\mu}^{15}/\sqrt{6}$
as diagonal bosons. These diagonal Hermitian generators will supply
the neutral gauge bosons $A_{\mu}^{em}$, $Z_{\mu}$, $Z_{\mu}^{\prime\prime}$and
$Z_{\mu}^{\prime\prime}$. Therefore, on the diagonal terms in eq.(\ref{Eq.6})
a generalized Weinberg transformation (gWt) must be performed in order
to consequently separate the massless electromagnetic field from the
other three neutral massive fields. One of the two massive neutral
fields is nothing but the $Z^{0}$-boson of the SM. The details of
the general procedure with gWt can be found in Ref. \cite{key-32}
and its concrete realization in the model of interest here in Refs.
\cite{key-19,key-25} where the neutral currents for all $Z_{\mu}$,
$Z_{\mu}^{\prime}$and $Z_{\mu}^{\prime\prime}$ are completely determined
and the boson mass spectrum as a function of the unique remaining
free parameter ($a$) is calculated.

\textbf{Scalar sector and spontaneous symmetry breaking}

In the general method \cite{key-32}, the scalar sector of any $SU(N)_{L}\otimes U(1)_{Y}$
electro-weak gauge model must consist of $n$ Higgs multiplets $\phi^{(1)}$,
$\phi^{(2)}$, ... , $\phi^{(n)}$ satisfying the orthogonal condition
$\phi^{(i)+}\phi^{(j)}=\varphi^{2}\delta_{ij}$ in order to eliminate
unwanted Goldstone bosons that could survive the SSB. Here $\varphi\sim\left(1,1,0\right)$
is a gauge-invariant real field acting as a norm in the scalar space
and $n$ is the dimension of the fundamental irreducible representation
of the gauge group. The parameter matrix $\eta=\left(\eta_{0},\eta{}_{1},\eta{}_{2}..,\eta{}_{n}\right)$
with the property $Tr\eta^{2}=1-\eta_{0}^{2}$ is a key ingredient
of the method: it is introduced in order to obtain a non-degenerate
boson mass spectrum. Obviously, $\eta_{0},\eta{}_{i}\in[0,1)$. Then,
the Higgs Ld reads:

\begin{equation}
\mathcal{L}_{H}=\frac{1}{2}\eta_{0}^{2}\partial_{\mu}\varphi\partial^{\mu}\varphi+\frac{1}{2}\sum_{i=1}^{n}\eta_{i}^{2}\left(D_{\mu}\phi^{(i)}\right)^{+}\left(D^{\mu}\phi^{(i)}\right)-V(\phi^{(i)})\label{Eq. 7}\end{equation}
 where $D_{\mu}\phi^{(i)}=\partial_{\mu}\phi^{(i)}-i(gA_{\mu}+g^{\prime}y^{(i)}A_{\mu}^{0})\phi^{(i)}$
act as covariant derivatives of the model. $g$ and $g^{\prime}$
are the coupling constants of the groups $SU(N)_{L}$ and $U(1)_{X}$
respectively. Real characters $y^{(i)}$ stand as a kind of hyper-charge
of the new theory.

For the particular 3-4-1 model under consideration here the most general
choice of parameters is given by the matrix

\begin{equation}
\eta^{2}=(1-\eta_{0}^{2})\mathrm{Diag}\left(\frac{1}{2}a-b,\frac{1}{2}a+b,c-a,1-c\right),\label{Eq.8}\end{equation}
where, for the moment, $a$,$b$ and $c$ are arbitrary non-vanishing
real parameters. Obviously, $\eta_{0},c\in[0,1)$, $a\in(0,c)$ and
$b\in(-a,+a)$. It obviously meets the trace condition required by
the general method for any $a,b\in[0,1)$. After imposing the phenomenological
condition $M_{Z}^{2}=M_{W}^{2}/\cos^{2}\theta_{W}$ (confirmed at
the SM level) the procedure of diagonalizing the neutral boson mass
matrix \cite{key-25} reduces to one the number of parameters, so
that the parameter matrix reads finally

\begin{equation}
\eta^{2}=(1-\eta_{0}^{2})\mathrm{Diag}\left(\frac{a}{2}(1-\tan^{2}\theta_{W}),\frac{a}{2}(1+\tan^{2}\theta_{W}),\frac{1-a}{2},\frac{1-a}{2}\right),\label{Eq.9}\end{equation}
while the 4 scalar 4-plets of the Higgs sector are represented by
$\phi^{(1)}\sim(\mathbf{1,4},-3/2)$ and $\phi^{(2)},\phi^{(3)},\phi^{(4)}\sim(\mathbf{1,4},1/2)$. 

With the following content in the scalar sector of the 3-4-1 model
of interest here and based on the redefinition of the scalar quadruplets
from the general method as $\eta{}_{1}\phi^{(1)}\rightarrow\chi$,
$\eta{}_{2}\phi^{(2)}\rightarrow\rho$, $\eta{}_{3}\phi^{(3)}\rightarrow\zeta$
and $\eta{}_{4}\phi^{(4)}\rightarrow\xi$

\begin{equation}
\left(\begin{array}{c}
\chi_{1}^{0}\\
\chi_{2}^{-}\\
\chi_{3}^{-}\\
\chi_{4}^{-}\end{array}\right)\sim(\mathbf{1},\mathbf{4},-3/2)\,\left(\begin{array}{c}
\rho_{1}^{+}\\
\rho_{2}^{0}\\
\rho_{3}^{0}\\
\rho_{4}^{0}\end{array}\right),\;\left(\begin{array}{c}
\zeta_{1}^{+}\\
\zeta_{2}^{0}\\
\zeta_{3}^{0}\\
\zeta_{4}^{0}\end{array}\right),\;\left(\begin{array}{c}
\xi_{1}^{+}\\
\xi_{2}^{0}\\
\xi_{3}^{0}\\
\xi_{4}^{0}\end{array}\right)\sim(\mathbf{1},\mathbf{4},1/2)\,,.\label{Eq. 10}\end{equation}
 one can achieve via the SSB the following vacuum expectation values
(VEV) in the unitary gauge:

\[
\left(\begin{array}{c}
\eta_{1}\left\langle \varphi\right\rangle +H_{\chi}\\
0\\
0\\
0\end{array}\right),\left(\begin{array}{c}
0\\
\eta_{2}\left\langle \varphi\right\rangle +H_{\rho}\\
0\\
0\end{array}\right),\left(\begin{array}{c}
0\\
0\\
\eta_{3}\left\langle \varphi\right\rangle +H_{\zeta}\\
0\end{array}\right),\left(\begin{array}{c}
0\\
0\\
0\\
\eta_{4}\left\langle \varphi\right\rangle +H_{\xi}\end{array}\right)\]
or equivalently

\[
\frac{1}{\sqrt{2}}\left(\begin{array}{c}
\sqrt{a(1-\tan^{2}\theta_{W})}\left\langle \varphi\right\rangle +H_{\chi}\\
0\\
0\\
0\end{array}\right),\;\frac{1}{\sqrt{2}}\left(\begin{array}{c}
0\\
\sqrt{a(1-\tan^{2}\theta_{W})}\left\langle \varphi\right\rangle +H_{\rho}\\
0\\
0\end{array}\right),\]

\begin{equation}
\frac{1}{\sqrt{2}}\left(\begin{array}{c}
0\\
0\\
\sqrt{(1-a)}\left\langle \varphi\right\rangle +H_{\zeta}\\
0\end{array}\right),\quad\quad\frac{1}{\sqrt{2}}\left(\begin{array}{c}
0\\
0\\
0\\
\sqrt{(1-a)}\left\langle \varphi\right\rangle +H_{\xi}\end{array}\right)\label{Eq. 11}\end{equation}
.with the overall VEV

\[
\left\langle \varphi\right\rangle =\frac{\sqrt{\mu_{1}^{2}\eta_{1}^{2}+\mu_{2}^{2}\eta_{2}^{2}+\mu_{3}^{2}\eta_{3}^{2}+\mu_{4}^{2}\eta_{4}^{2}}}{\sqrt{2\left(\lambda_{1}\eta_{1}^{4}+\lambda_{2}\eta_{2}^{4}+\lambda_{3}\eta_{3}^{4}+\lambda_{4}\eta_{4}^{4}\right)+\lambda_{5}\eta_{1}^{2}\eta_{2}^{2}+\ldots+\lambda_{10}\eta_{3}^{2}\eta_{4}^{2}}}\]
resulting from the minimum condition applied to the potential

\begin{equation}
\begin{array}{ccl}
V & = & -\mu_{1}^{2}\chi^{\dagger}\chi-\mu_{2}^{2}\rho^{\dagger}\rho-\mu_{3}^{2}\zeta^{\dagger}\zeta-\mu_{4}^{2}\xi^{\dagger}\xi\\
\\ &  & +\lambda_{1}\left(\chi^{\dagger}\chi\right)^{2}+\lambda_{2}\left(\rho^{\dagger}\rho\right)^{2}+\lambda_{3}\left(\zeta^{\dagger}\zeta\right)^{2}+\lambda_{4}\left(\xi^{\dagger}\xi\right)^{2}\\
\\ &  & +\lambda_{5}\left(\chi^{\dagger}\chi\right)\left(\rho^{\dagger}\rho\right)+\lambda_{6}\left(\chi^{\dagger}\chi\right)\left(\zeta^{\dagger}\zeta\right)+\lambda_{7}\left(\chi^{\dagger}\chi\right)\left(\xi^{\dagger}\xi\right)\\
\\ &  & +\lambda_{8}\left(\rho^{\dagger}\rho\right)\left(\zeta^{\dagger}\zeta\right)+\lambda_{9}\left(\rho^{\dagger}\rho\right)\left(\xi^{\dagger}\xi\right)+\lambda_{10}\left(\zeta^{\dagger}\zeta\right)\left(\xi^{\dagger}\xi\right)\end{array}\label{Eq. 12}\end{equation}

The above potential has the simplest form allowed by both the gauge
symmetry and the restrictive orthogonal condition in the scalar sector.
The scales it provides are, obviously, splitted when parameter tuning
favors $a\rightarrow0,$ such as $v\simeq v^{\prime}\sim\sqrt{a}$
(responsible for the SM phenomenology) and $V=V^{\prime}\sim\sqrt{1-a}$
(responsible for the heavier degrees of freedom in the 3-4-1 model)
respectively. A more detailed redefinition (with worked out consequences
for the Higgs sector) can be observed in the case of the 3-3-1 model
(see Ref.\cite{key-37}), but for our purposes here it needs no further
development.

\section{Implementing seesaw mechanism}

With the above ingredients one can construct the Yukawa Ld allowed
by the gauge symmetry in the 3-4-1 model. It simply give rise to the
following terms:

\begin{equation}
\begin{array}{ll}
\mathcal{-L}_{Y}^{\nu} & =h_{\rho}\overline{L}\rho^{\dagger}N_{R}+h_{\eta}\overline{L}\zeta^{\dagger}N_{R}+h_{\xi}\overline{L}\xi^{\dagger}N_{R}+\frac{1}{2}h_{\varphi}\overline{N_{R}^{c}}\eta_{0}\varphi N_{R}\\
\\ & +\frac{1}{2\Lambda}h^{m}\varepsilon^{ijk}\left(\overline{L}\right)_{i}\left(L^{c}\right)_{j}\left(\rho_{k}^{\dagger}\zeta_{m}^{\dagger}+\zeta_{k}^{\dagger}\xi_{m}^{\dagger}+\xi_{k}^{\dagger}\rho_{m}^{\dagger}\right)+h.c.\end{array}\label{Eq. 13}\end{equation}
where all $h$s are $3\times3$ complex Yukawa matrices, their lower
index indicating the particular Higgs each one connects with. 

The Yukawa Ld leads straightforwardly to the following neutrino mass
terms:

\begin{equation}
\begin{array}{ll}
\mathcal{-L}_{mass}^{\nu} & =h_{\rho}\overline{\nu_{L}}N_{R}\left\langle \rho\right\rangle +h_{\zeta}\overline{N_{L}^{\prime}}N_{R}\left\langle \zeta\right\rangle +h_{\xi}\overline{N_{L}^{\prime\prime}}N_{R}\left\langle \xi\right\rangle \\
\\ & +\frac{1}{2}h_{\varphi}\overline{N_{R}}N_{R}^{c}\eta_{0}\left\langle \varphi\right\rangle +\frac{1}{2\Lambda}\left(h^{\prime}-h^{\prime T}\right)\overline{\nu_{L}}N_{L}^{c}\left(\left\langle \zeta\right\rangle \left\langle \xi\right\rangle +\ldots\right)\end{array}\label{Eq. 14}\end{equation}

Evidently, $\Lambda$ is the cut-off scale of the model, up to which
it remains renormalizable as an effective theory. The Yukawa terms
allow one to construct the quasi-inverse seesaw mechanism by displaying
them into the following $9\times9$ complex matrix:

\begin{equation}
M=\left(\begin{array}{ccccc}
0 &  & \frac{h}{\Lambda}\left(1-a\right) &  & h_{\rho}\sqrt{a\left(1+\tan^{2}\theta_{W}\right)}\\
\\\frac{h^{T}}{\Lambda}\left(1-a\right) &  & 0 &  & H\sqrt{1-a}\\
\\h_{\rho}^{T}\sqrt{a\left(1+\tan^{2}\theta_{W}\right)} &  & H^{T}\sqrt{1-a} &  & \frac{1}{2}h_{\varphi}\eta_{0}\end{array}\right)\left\langle \varphi\right\rangle \label{Eq. 15}\end{equation}
where $h=h^{\prime}-h^{\prime T}$. Since $\left\langle \zeta\right\rangle =\left\langle \xi\right\rangle $,
one can construct the Dirac mass term $H\overline{N_{L}}N_{R}\sqrt{1-a}\left\langle \varphi\right\rangle $.
The new state $N_{L}$ can be taken either $N_{L}^{\prime}$ or $N_{L}^{\prime\prime}$,
as both exhibit the same quantum numbers and couplings to neutral
currents (see couplings' Table in Ref. \cite{key-25}). Therefore
one can safely consider that $N_{L}^{\prime}=N_{L}^{\prime\prime}=N_{L}$.

Due to the non-zero $h_{\rho}$ this matrix is slightly different
from the traditional inverse seesaw mechanism \cite{key-38} - \cite{key-40},
but its resulting effects - we prove in the following - are phenomenologically
plausible. However, this kind of seesaw matrix appears in the literature,
see for instance Refs. \cite{key-41,key-42}. This $9\times9$ complex
matrix can be displayed as:

\begin{equation}
M=\left(\begin{array}{ccc}
0 &  & m_{D}\\
\\m_{D}^{T} &  & M_{N}\end{array}\right)\label{Eq. 16}\end{equation}
with $m_{D}=\left(\begin{array}{ccc}
\frac{h}{\Lambda}\left(1-a\right) &  & h_{\rho}\sqrt{a\left(1+\tan^{2}\theta_{W}\right)}\end{array}\right)$ a $3\times6$ complex matrix and $M_{N}=\frac{1}{2}\left(\begin{array}{cc}
0 & H\sqrt{2\left(1-a\right)}\\
H^{T}\sqrt{2\left(1-a\right)} & h_{\varphi}\eta_{0}\end{array}\right)$ a $6\times6$ complex matrix acting in the seesaw formula.

By diagonalizing the above matrix one gets the physical neutrino matrices
as: $M\left(\nu_{L}\right)\simeq-m_{D}\left(M_{N}^{-1}\right)m_{D}^{T}$
and $M\left(\nu_{R},N_{R}\right)\simeq M_{N}$ which yield:

\begin{equation}
\begin{array}{cl}
M\left(\nu_{L}\right) & \simeq-\frac{\left(1-a\right)\eta_{0}\left\langle \varphi\right\rangle }{\Lambda^{2}}\left[h\left(H^{-1}\right)^{T}\left(h_{\varphi}\right)\left(H^{-1}\right)h^{T}\right]\\
\\ & +\frac{\sqrt{a\left(1-a\right)\left(1+\tan^{2}\theta_{W}\right)}\left\langle \varphi\right\rangle }{\Lambda}\left[\left(h_{\rho}\right)\left(H^{-1}\right)h^{T}+h\left(H^{-1}\right)^{T}\left(h_{\rho}\right)^{T}\right]\end{array}\label{Eq. 17}\end{equation}

\begin{equation}
\left(\begin{array}{cc}
M\left(\nu_{R}\right) & 0\\
\\0 & M\left(N_{R}\right)\end{array}\right)=\left(\begin{array}{cc}
H\sqrt{1-a}+\frac{1}{2}h_{\varphi}\eta_{0} & 0\\
\\0 & -H\sqrt{1-a}+\frac{1}{2}h_{\varphi}\eta_{0}\end{array}\right)\left\langle \varphi\right\rangle \label{Eq. 18}\end{equation}
The first term in Eq.(\ref{Eq. 17}) is obviously much smaller than
the second one due to $\sim1/\Lambda^{2}$. 

Now, for the sake of simplicity and to get a rapid estimation of the
masses, one can suppose without loosing the generality, the proportionality
\begin{equation}
h_{\rho}=\alpha H\label{Eq. 19}\end{equation}
Consequently, one gets the left-handed neutrino mass matrix as the
complex $3\times3$ matrix:

\begin{equation}
M\left(\nu_{L}\right)\simeq\alpha\frac{\sqrt{a\left(1-a\right)\left(1+\tan^{2}\theta_{W}\right)}\left\langle \varphi\right\rangle }{\Lambda}\left(h^{T}+h\right)\label{Eq. 20}\end{equation}

It is evident that it is a pure Majorana mass matrix since $M\left(\nu_{L}\right){}^{T}=M\left(\nu_{L}\right)$
holds. Assuming a natural order of magnitude, say $h_{\rho},h,h_{\varphi}\sim O(1)$,
one can estimate the order of magnitude of the individual masses in
this matrix as

\begin{equation}
TrM\left(\nu_{L}\right)\simeq6\alpha\frac{\sqrt{\left(1+\tan^{2}\theta_{W}\right)}\left\langle \varphi\right\rangle _{SM}}{\Lambda\sqrt{1-\eta_{0}^{2}}}\label{Eq. 21}\end{equation}
since $m\left(W\right)=\frac{1}{2}g\left\langle \varphi\right\rangle _{SM}=\frac{1}{2}g\sqrt{\left(1-\eta_{0}^{2}\right)a}\left\langle \varphi\right\rangle $
(see \cite{key-25}) .

The right-handed neutrinos acquire some degenerate masses \begin{equation}
M\left(\nu_{R}\right)\simeq M\left(N_{R}\right)\simeq\frac{1}{2}h_{\varphi}\eta_{0}\left\langle \varphi\right\rangle \label{Eq. 22}\end{equation}
if $H\sim O(10^{-x})$, $\eta_{0}\sim O(10^{-y})$ and $x\gg y$.
Otherwise, for $x=y$, the exotic right-handed $N_{R}$s could come
out as very tiny or even massless, while for the usual right-handed
neutrinos $M\left(\nu_{R}\right)\simeq H\left\langle \varphi\right\rangle \simeq\alpha^{-1}h_{\rho}\left\langle \varphi\right\rangle $.

\section{Tuning the parameters}

Now one can tune the parameters in this particular 3-4-1 model in
order to get phenomenologically viable predictions. Obviously, both
$a$, $\eta_{0}\in(0,1)$. Since $\eta_{0}$ is the parameter responsible
with the lepton number violation, one can keep it very small, say
$\eta_{0}\sim10^{-6}$ (or even smaller) in order to safely consider
that the global $U(1)_{leptonic}$ symmetry is very softly (quite
negligible) violated by the Majorana coupling it introduces.

When comparing the boson mass spectrum in this model - obtained both
by using the general Cot\u{a}escu method \cite{key-32} and the SM
calculations \cite{key-1} - one gets a scales connection:

\begin{equation}
\sqrt{\left(1-\eta_{0}^{2}\right)a}=\frac{\left\langle \varphi\right\rangle _{SM}}{\left\langle \varphi\right\rangle }\label{Eq. 23}\end{equation}

It becomes obviously that $\eta_{0}$ has no part to play in the breaking
scales splitting. The later is determined quite exclusively by $a$.
If one takes $\left\langle \varphi\right\rangle _{SM}\simeq246$GeV
and $\left\langle \varphi\right\rangle \simeq1-10$ TeV then $a\simeq\left(0.0006\;,\;0.06\right)$.

With these plausible settings the individual neutrino masses come
out in the subsequent hierarchy:

\begin{equation}
\sum m\left(\nu_{L}\right)\simeq10^{12}\left(\frac{\alpha}{\Lambda}\right)eV\label{Eq. 24}\end{equation}

\begin{equation}
\sum m\left(\nu_{R}\right)\simeq10^{-7}\left\langle \varphi\right\rangle \label{Eq. 25}\end{equation}
where $\left(\frac{\alpha}{\Lambda}\right)^{2}\left(1-a\right)\eta_{0}\left\langle \varphi\right\rangle $
- as the first term in Eq.(\ref{Eq. 17}) - is negligible under the
restriction $\left(\frac{\alpha}{\Lambda}\right)^{2}\sim10^{-24}$
(from Eq.(\ref{Eq. 24})) and $\left\langle \varphi\right\rangle \sim10^{12}$eV. 

One can observe from Eqs. (\ref{Eq. 24}) - (\ref{Eq. 25}) in the
case with massless exotic $N_{R}$s, the smaller the masses of the
right-handed $\left(\nu_{R}=N_{L}^{c}\right)$ neutrinos, the higher
the cut-off of the effective theory in this particular 3-4-1 model.
That means, if $m\left(\nu_{R}\right)\sim10^{-3}$eV then the effective
theory is valid up to $\Lambda\sim10^{15}$GeV which is the GUT scale.
However this can be kept valid up to such high energies even though
the right-handed massive neutrinos lie at any level in the sub-TeV
region at the expense of assuming six quasi-degenerate such right-handed
neutrinos. 

Furthermore, one can enforce some extra flavor symmetries in the lepton
sector in order to dynamically get the appropriate PMNS mixing matrix.
Some discrete groups, such as $A_{4}$\cite{key-47,key-48}, $S_{4}$\cite{key-49}
or $S_{3}$\cite{key-50,key-51} were employed in 3-3-1, and the procedure
can be similarly applied in order to accomplish this task in the model
of interest here.

\section{Concluding remarks}

We discussed in this brief letter the possible realization of a quasi-inverse
seesaw mechanism in a particular 3-4-1 gauge model with ''lepton
number violating'' exotic Majorana neutrinos added. The Cot\u{a}escu
general method of treating gauge models with high-symmetries is employed
and works as a suitable framework for such a purpose. It successfully
provides us not only with the one-parameter mass spectrum but also
with the lepton number violating terms needed for a plausible inverse
seesaw mechanism, due to the possibility of coupling the scalar $\varphi$
to exotic Majorana neutrinos. To the extent of our knowledge, in low
energy models one finds no such terms to give masses to exotic neutrinos,
so that some extra assumptions (usually from GUT theories) are invoked.
These two characteristics single out our approach from other recent
similar attempts (for instance, in 3-3-1 models see Refs. \cite{key-52}
- \cite{key-58}). The details of the mixing in the neutrino sector
are closely related to the entries in $h$, $h_{\rho}$ and $h_{\varphi}$
but this lies beyond the scope of this letter and will be presented
elsewhere. 

Such SM-extensions are appealing for they proves themselves able to
recover all the results of the SM and in addition exhibit a lot of
assets: they require precisely 3 fermion generations, their algebraic
structures dictate the observed charge quantization, they can predict
a testable Higgs phenomenology and, as we presented here, are utterly
promising for neutrino phenomenology.

\end{document}